\documentclass{article}

\newcommand{\ket}[1]{{|#1\rangle}}

\author{Christof Zalka \\
Department of Combinatorics \& Optimization \\
University of Waterloo \\
Waterloo, Ontario \\
Canada N2L 3G1}

\title{Introduction to Quantum Computers and Quantum Algorithms}

\begin{document}
\maketitle

\begin{abstract}
This is a short introduction to Quantum Computing intended for
physicists. The basic idea of a quantum computer is introduced. Then
we concentrate on Shor's integer factoring algorithm.
\end{abstract}

       
\section{What is a Quantum Computer?}
In this first lesson I want to introduce the basics necessary to
understand quantum algorithms: What a quantum computer is, quantum
gates and how to do usual computations on a quantum computer, namely
``reversible computation".

\subsection{A Quantum Computer = many 2-level systems (qubits)}
A ``qubit" (or ``quantum bit") is simply a quantum mechanical 2 level
system. In theoretical considerations, for convenience we usually
assume that the 2 levels are energy degenerate. So a qubit is any
quantum mechanical system with a 2 dimensional Hilbert space, or a
system of which we use only a 2 dimensional subspace. Examples are the
spin degree of freedom of a spin $1/2$ particle, the polarization of a
single photon or an atom or ion in the ground state or a particular
excited state. Also we can use the ground- and first excited state of
a harmonic oscillator or of a mode of the electromagnetic field, thus
no photon (vacuum) or 1 photon. In the 2 dimensional Hilbert space we
choose 2 orthonormal basis states and denote them by $\ket{0}$ and
$\ket{1}$. (For a non energy-degenerate qubit these would usually be
the energy eigenstates, like the electronic ground state and some
excited state of an atom.)

A quantum computer simply consists of many such qubits. For
convenience we imagine that usually they do not interact, thus when we
don't intervene, the time evolution of the quantum computer is trivial
(say $H_0 =0$). Now the Hilbert space of many qubits is the tensor
product of their Hilbert spaces. A natural basis is given by states of
the form

$$\ket{1101001 \dots} \doteq
\ket{1}\ket{1}\ket{0}\ket{1}\ket{0}\ket{0}\ket{1}\dots $$  
Thus the overall Hilbert space is spanned by all binary strings of
some length (corresponding to each qubit being either in state
$\ket{0}$ or in state $\ket{1}$). Note that the dimension of this
space increases exponentially with the number of qubits, and so the
number of probability amplitudes necessary to specify a state
increases exponentially.

As with conventional computers, we often imagine that bit strings
stand for numbers in binary, so we can label such basis states with
numbers. Then a (pure) state of a $n$ qubit quantum computer (QC) can
be written

$$\ket{\Psi_{QC}} = \sum_{x=0}^{2^n-1} a_x \ket{x}$$
with the normalization $\sum_x |a_x|^2 =1$.

{\bf exercise:} Estimate the maximal number of qubits such that it is
still possible to store the amplitudes describing their overall state
on some presently available computer. What about a computer using all
matter in the visible universe?

\subsection{Computing: ``quantum gates"}
Usually (and without loss of generality) we assume that the quantum
computer is initially in the ``all 0" state, thus the product state
where all qubits are in state $\ket{0}$:

$$\ket{QC_{ini}} = \ket{0000 \dots}$$
The computation consists of a sequence of unitary operations that are
realistically assumed to only act on few qubits at a time. Formally,
when we apply e.g. some $U(8)$ transformation to 3 qubits, the overall
transformation is the tensor product of this transformation with the
identity on the remaining qubits. Clearly, it is necessary to apply
unitary transformations to more than just individual qubits in order
to arrive at interesting states, as otherwise we will be stuck with
(unentangled) product states. [{\bf exercise:} Check that typically
all amplitudes describing the state of the quantum computer are
changed when a quantum gate (a unitary operation) is applied to, say,
one qubit.]

It turns out that it is enough if we are able to apply unitary
transformations (= ``quantum gates") to any two qubits (thus $U(4)$),
and of course also to individual qubits ($U(2)$). In the following it
will also be convenient to assume that we can directly do some 3 qubit
quantum gate, although it could really be decomposed into 1- and 2
qubit gates.

Below is a graphical representation of a quantum computation on 5
qubits (which is little for theoreticians, but much for
experimentalists...). The single dots represent some $U(2)$
transformations on single qubits while the pairs of dots joined by a
vertical line stand for 2-qubit gates, thus some $U(4)$s. Time runs
from left to right. Take care not to confuse the qubits (the 5
horizontal lines) with the 32 basis states.

\newcommand{\NOT}{\begin{picture}(10,10)
    \put(0,0){\circle{10}} \put(0,-5){\line(0,1){10}}
    \put(-5,0){\line(1,0){10}} \end{picture}}
\newcommand{\C}{\circle*{5}}

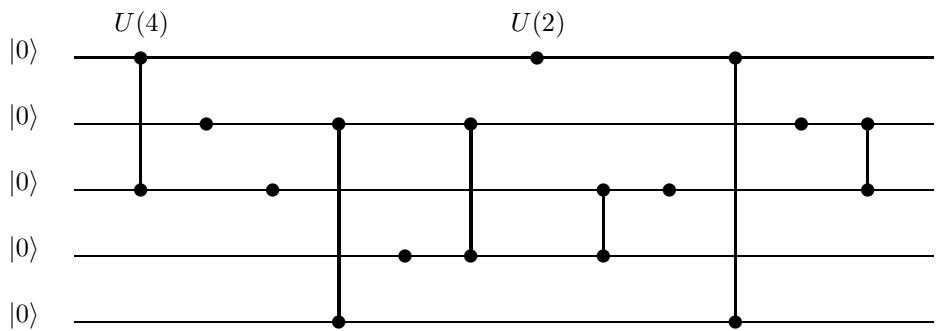
\begin{figure}
\begin{center}
\begin{picture}(300,145)

\thinlines \multiput(25,0)(0,25){5}{\line(1,0){325}}
\multiput(0,0)(0,25){5}{\put(0,0){$\ket{0}$}} \thicklines

\put(40,110){$U(4)$} \put(190,110){$U(2)$}

\put(50,0){ \put(0,50){\line(0,1){50}} \put(0,100){\C} \put(0,50){\C}}
\put(75,75){\C}
\put(100,50){\C}
\put(125,0){ \put(0,0){\line(0,1){75}} \put(0,0){\C} \put(0,75){\C}}
\put(150,25){\C}
\put(175,0){ \put(0,25){\line(0,1){50}} \put(0,25){\C} \put(0,75){\C}}
\put(200,100){\C}
\put(225,0){ \put(0,25){\line(0,1){25}} \put(0,25){\C} \put(0,50){\C}}
\put(250,50){\C}
\put(275,0){ \put(0,0){\line(0,1){100}} \put(0,0){\C} \put(0,100){\C}}
\put(300,75){\C}
\put(325,0){ \put(0,50){\line(0,1){25}} \put(0,50){\C} \put(0,75){\C}}
\end{picture}
\end{center}
\vspace*{1cm}
\caption{A sequence of quantum gates on 5 qubits, all initialized in
state $\ket{0}$. Time runs from left to right. Only 1-qubit and
2-qubit gates are used, as this is sufficient for universal quantum
computation. A complete quantum computation would also include a
measurement of each qubit at the end (on the right).}
\label{gates}
\end{figure}

\subsubsection{How ``quantum gates" are carried out}
Quantum gates are done by ``switching on" terms in the Hamilton
operator for an appropriate time. Physically, this is done by applying
``exterior fields" to the qubits, thus fields which can be treated as
classical. So called ``coherent states" as e.g. produced by a laser or
radio transmitter are like this, thus they behave nearly
classically. It is remarkable that such states exists which allow us
to manipulate a quantum system without getting entangled with it, thus
without causing decoherence. Also it is not obvious that these states
occur naturally.

1-qubit gates are usually rather easy to carry out. In an ion trap
quantum computer, one can shine a laser with the right frequency for
some time at an ion. By tuning the frequency to the energy difference
between the ground- and excited states (thus resonance), we induce
``Rabi oscillations" between $\ket{0}$ and $\ket{1}$. Note that to
carry out the gate correctly we also have to control the intensity,
time and phase of the laser.

In most hardware proposals, 2-qubit gates are more difficult to carry
out. One possibility is to simply bring 2 qubits closely together so
that they interact.

\subsection{conventional computation on a quantum computer}
Most of Shor's algorithm is simply a conventional computation, but
applied to a superposition of (exponentially) many ``conventional"
states. For a computation to work on superpositions it has to be
coherent, thus is must consists of (unitary) quantum gates, excluding
e.g. measurements. Also ``throwing away" qubits would not be allowed,
as it would destroy coherence. But for Shor's algorithm it is
necessary that at the end only a certain function is calculated and
not also some intermediate results. In the following I will also show
how the unwanted intermediate results can be ``deleted" in a coherent
way.

\subsubsection{reversible gates}
First of all we want only to use gates which have as many output bits
as input bits (unlike e.g. an AND gate which is 2 $\to$ 1), as we
don't want to destroy qubits.

It is easy to imagine a classical computation with such gates: The
computer is a bit string of constant length. We can now act on
individual bits with, say, NOT or RESET (reset to 0). In addition, we
can pick any 2 bits in the bit string and e.g. leave the first bit
unchanged and set the second one to the AND of the 2 bits, or SWAP the
2 bits. Clearly in this way any computation can be done. (It is well
known that any classical computation can be done by using just NAND (=
not AND), actually there used to be chips with several NAND gates on
them which could then be wired into logic circuits by engineers.)

We want to do classical computation with unitary gates to preserve
coherence. The only unitary transformations which map all
``computational basis states" (binary strings) to such basis states
(and not superpositions), are permutations of these basis
states. (Thus in each row and column of the unitary matrix there is
one 1.) Such classical gates are called ``reversible", as they are 1
to 1. There is only one non-trivial such 1-qubit gate, the NOT:

$$U_{NOT}: \quad \ket{0} \to \ket{1} \qquad \ket{1} \to \ket{0}$$
An important 2-qubit gate is the ``controlled-NOT" (CNOT or XOR),
which doesn't change the first bit but changes the second one if the
first is a 1:

$$U_{CNOT}: \quad \ket{a,b} \to \ket{a,a ~XOR~ b} = \ket{a,a\oplus b}$$
(XOR means ``exclusive OR", thus either $a$ or $b$ but not both;
$\oplus$ is addition modulo 2) [{\bf exercise}: check that CNOT is a
permutation of the 4 possible 2-bit strings, and thus is unitary. Also
write down the 4x4 matrix representing the overall transformation when
we act on each of 2 qubits with a NOT (thus find the tensor product of
the two NOTs).]

It is not difficult to show that the above 2 gates are not enough to
do universal computation. We still lack something that allows to
compute an AND, but this is not possible with reversible 2-bit gates
[{\bf exercise:} show that if one of the output bits of a 2-bit gate
is the AND of the inputs, then the gate can't be reversible.] For this
we use the CCNOT (= ``Toffoli gate") which acts as follows on 3
(qu)bits (thus it's a $U(8)$):

$$U_{CCNOT}: \quad \ket{a,b,c} \to \ket{a,b,c ~XOR~ (a ~AND~ b)} =
\ket{a,b,c \oplus(a \cdot b)}$$ 
Thus the last bit is flipped only if $a$ and $b$ are both 1.

\subsection{``garbage uncomputing"}
It is now clear that starting from some input $\ket{x}$ (thus the
binary string representing $x$) we can compute any function
$f(x)$. But we will probably also produce some unwanted output
$g(x)$. E.g. if we compute the AND of two bits with the CCNOT, the two
original bits will still be around. Of course besides $\ket{x}$, we
will also need qubits initialized to $\ket{0}$. In summary, what we
can do is:

$$\ket{x,0} \to \ket{f(x),g(x)}$$
where the $0$ stands for several qubits. In the following, I will not
always indicate when in the process of computing we will add such
``fresh" qubits in state $\ket{0}$.

As they are unitary, all the gates we use are reversible; thus, we can
also undo the whole computation by applying the inverse gates in
reversed order. The trick to get rid of the unwanted ``garbage" $g(x)$
now is to first copy the wanted $f(x)$ into a ``save place" and then
undo the original computation (which doesn't touch the ``save
place''). Copying into a ``save place" can simply be done by ``XORing"
each bit of $f(x)$ into a fresh bit in state $0$. Thus overall:

$$\ket{x,0,0} ~\stackrel{U_f}{\to}~ \ket{f(x),g(x),0} ~\stackrel{copy~
f(x)}{\to}~ \ket{f(x),g(x),f(x)} ~\stackrel{U_f^{-1}}{\to}~
\ket{x,0,f(x)} $$ 
Thus we will still have $x$ in the final state, but this is
unavoidable if $f(x)$ is not bijective (1 to 1), as otherwise the
overall computation would not be unitary. The ``work qubits", which at
the end of the computation are again in state $\ket{0}$, pose no
problem, as they are now again unentangled with the other qubits and
could thus safely be thrown away.

{\bf exercise:} If $f(x)$ is 1 to 1 and if we know an (efficient)
algorithm for its inverse, we can do $\ket{x} \to \ket{f(x)}$ by using
similar tricks as above. Hint: use the reverse of the algorithm which
computes $f^{-1}$, thus $(U_{f^{-1}})^{-1}$.

\subsection{measurement of the quantum computer}
At the end of a computation, the quantum computer is observed and thus
its state ``collapses" to some binary string that will be the
(classical) output of the computation. Thus we assume that we measure
an observable whose (non-degenerate) eigenstates are the computational
basis states. This corresponds to measuring each qubit separately such
that it gets projected to either $\ket{0}$ or $\ket{1}$. It would be
unrealistic to assume that we can simply measure any observable on the
whole quantum computer. On the other hand, if we would like to make a
measurement corresponding to another basis, this is equivalent to
first doing some unitary transformation (a sequence of quantum gates)
and then doing the usual measurement. In this way, one can
e.g. effectively measure two qubits in the (maximally entangled)
``Bell basis".

\subsection{summary}
We have shown that, given a conventional algorithm to compute $f(x)$,
we can construct a sequence of quantum gates which act on basis states
as $\ket{x} \to \ket{x,f(x)}$. Because time evolution in quantum
theory is linear (unitary maps are linear), we can apply this to
superpositions and will get a superposition of the outputs:

$$\sum_x a_x~ \ket{x} ~\to~ \sum_x a_x~ \ket{x,f(x)}$$
Thus in a way we can compute the function $f(x)$ in one go in
``quantum parallelism" for many inputs $x$. Of course simply observing
the final superposition will simply give us a basis state at random
(with probability $|a_x|^2$), thus in Shor's algorithm we first do
something else before measuring.

\section{Shor's quantum factoring algorithm}
In 1994 Peter Shor found his famous algorithm which created much
interest in quantum computing. Probably this algorithm is still the
most important quantum algorithm. (In other words, there has not been
as much progress as many people had hoped for.)

\subsection{The problem: finding prim factors}

The problem is to write a given positive integer as a product of prim
numbers, e.g. $12827=101 \times 127$. It is not so difficult to find
small prime factors, but if there are several large prime factors, no
fast classical algorithm for factoring is known. Much better
algorithms are known than just trying to divide a number by all
possible prim factors (up to the square root). But the presently best
algorithm still takes time which is roughly exponential in the third
root of the number of digits: the number of elementary operations to
factor a number $N$ is roughly $10^{4 (\log_{10} N)^{1/3}}$. The
largest numbers that can presently be factored (with much computer
power and time) have about 150 decimal digits.

\subsubsection{public key cryptography: RSA}
Factoring is important because it would allow to break the important
``public key" cryptosystem RSA. With this system, it is possible to
make public how to encrypt messages, but this will not allow the
public to decrypt messages. (In principle of course, decryption is
possible, but it is thought that it would take too long.) Simplifying
things a bit, in RSA the public key needed to encrypt, is the product
$p \times q$ of two very large prim numbers, while the ``private key"
e.g. allowing a military headquarter to decrypt messages coming by
radio from the field, are the two large prim numbers $p$ and $q$
separately. So factoring $pq$ would allow breaking the code.

It is possible to calculate rather quickly whether a given number is a
prim number or not, so a computer can easily generate some large prim
numbers $p$ and $q$. (So it is also easy to find that a number is
composite (not prim), but this is not the same as actually finding the
factors.)

\subsection{Fermat's little theorem and its generalization}
Shor's algorithm relies on the fact that factoring can be reduced to
finding the period $r$ of some periodic function from the integers to
the positive integers: $f: Z \to N$, with $f(x+r)= f(x)$ for all $x$.

First consider Fermat's little theorem which says that $a^{p-1} ~mod~
p =1$, for a prim $p$ and any non-zero integer $a \in Z_p$ (thus $0 <
a < p$). The reason is simply that those $p-1$ numbers form the
multiplicative group $mod~ p$. However, in any finite group an element
to the power order of the group (number of elements) is the neutral
element: $g^{|G|} =e$.

The same can be done for the ring $mod~ pq$. In this ring all numbers
$< pq$ which are coprim to $pq$ (thus their greatest common divisor
(gcd) is 1) have a multiplicative inverse. There are $(p-1)(q-1)$ such
numbers, so this is the order of the multiplicative group.

[{\bf exercise:} Show that $mod~ n$ a number $m$ which is coprim to
$n$ has a multiplicative inverse, thus there is an integer $m'$ such
that $m' m=k n+1$ for some integer $k$. Hint: Euclid's
(non-quantum...) algorithm to find gcd(m,n) will also give $k$ and
$m'$.]

Thus we have for any $a$ coprim to $pq$ that $a^{(p-1)(q-1)} ~mod~ pq
=1$. Now consider the function $f_a(x)=a^x ~mod~ pq$. It is periodic
with period $r$ where $r$ is the smallest number such that $a^r ~mod~
pq =1$. In addition, within a period it is 1 to 1, thus

$$f(x) = f(y) \qquad \Longleftrightarrow~ \qquad x-y=k r \qquad
\mbox{($r$ = period, $k$ integer)}$$ 
In Shor's algorithm the period $r$ of $f(x)$ is determined. It is
clear that $r$ has to be a fraction of $(p-1)(q-1)$, and usually it is
not a very small fraction. If we knew $(p-1)(q-1)$ (and of course
$pq$) we could easily find $p$ and $q$ by solving a quadratic
equation. I hope it is therefore plausible that finding $r$ also
essentially solves the problem. I will not show the actual way Shor
proposed to get the prim factors after finding $r$. Also for
simplicity let us imagine that, as in RSA, we know that we have a
product of just 2 primes, although Shor's algorithm works for
factoring any number.

\subsection{Shor's algorithm}
Shor's algorithm starts by preparing the ``uniform amplitude
superposition" (or ``equally weighted superposition") for some number
$n$ of qubits:

$$\frac{1}{\sqrt{2^n}} \sum_{x=0}^{2^n-1} \ket{x}$$

Note that in this state every qubit is in the state $1/\sqrt{2}~
(\ket{0}+\ket{1})$, thus this is an unentangled state which can easily
be obtained.

In the main part of the algorithm we carry out the (really classical)
algorithm to compute the function $f(x)$ (thereby also adding some
``fresh" qubits, and doing the ``garbage uncomputing" as described
above):

$$\sum_x \ket{x} \to \sum_x \ket{x,a^x ~mod~ pq}$$
Above and in the following I will often leave away the state normalizations.

[{\bf exercise:} Show how to quickly compute ``modular exponentiation"
even with big numbers. Compute (it's possible by hand) $8^{65} ~mod~
37$. Clearly first computing $8^{65}$ and only then doing $mod~ 37$ is
not the best. With what power of the number of digits of the three
numbers involved (say all have $n$ digits) does the computation time
grow?]

Note: for Shor's algorithm it is also helpful that the necessary
``garbage uncomputing" can be done periodically and not only at the
end, as this would use a lot of work space (qubits).

\subsubsection{period finding}
\begin{figure}
\begin{center}
\begin{picture}(300,80)

\thinlines \put(-5,0){\line(1,0){270}}
\multiput(16,0)(28,0){9}{\line(0,1){63}}

\put(0,0){\line(0,-1){5}} \put(-3,-15){0}
\put(252,0){\line(0,-1){5}} \put(240,-15){$2^6-1$}

\put(16,0){\line(0,-1){5}} \put(10,-15){$x_0$}
\put(44,0){\line(0,-1){5}} \put(28,-15){$x_0+r$}
\put(72,0){\line(0,-1){5}} \put(62,-15){$x_0+2 r \qquad \dots$}

\end{picture}
\end{center}
\vspace*{1cm}
\caption{Amplitudes of a ``periodic" state of a 6 qubit register, as
it is obtained in Shor's algorithm: $\sum_k \ket{x_0+k \cdot r}$. (For
the picture I chose $x_0=4$ and $r=7$, so $k=0\dots 8$.)}
\label{periodic}
\end{figure}
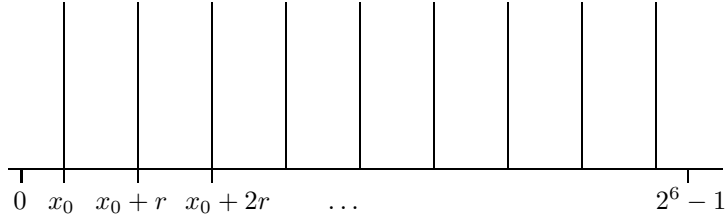

Let's now imagine that we measure the ``quantum register" in which
$f(x)$ is (this is actually not necessary, but makes the presentation
easier):

$$\sum_x \ket{x,f(x)} ~~\to~~ \sum_k \ket{x_0+k r,f(x_0)}$$
Where we assume that we chose the range $0\dots 2^n-1$ of the input
$x$ to cover many periods $r$ (see later). Thus we will observe at
random some value $f(x_0)$ and collapse the state of the QC to the
corresponding subspace. Because of the periodicity of $f(x)$ many
input values $x$ will give the same value, thus with the collapse we
will still have a superposition in the input register. As it is now
unentangled with the ``output register" (which is in a fixed basis
state), we can only look at the input register:

$$\sum_k \ket{x_0+k r}$$
where $k$ runs from $0$ to about $2^n/r$. If we plot the amplitudes as
a function of the number of the basis state, we get equally spaced
peaks, beginning not at 0 but with an offset $x_0$ (see figure
\ref{periodic}). Observing this superposition will not help in finding
the period $r$, as in each run of the algorithm we will get a
different random offset.

\subsubsection{The Quantum (Fast) Fourier Transform}
The Quantum (Fast) Fourier Transform applies the discrete Fourier
Transform to the amplitudes of a quantum register, thus:

$$QFFT: \quad \sum_{x=0}^{2^n-1} a_x \ket{x} \to \sum_{x=0}^{2^n-1}
\tilde a_x \ket{x} \qquad \mbox{with} \quad  \tilde a_x = 2^{-n/2}
\sum_{y=0}^{2^n-1} e^{2 \pi i x y /2^n} a_y $$ 
Note that this is a unitary transformation. It can be carried out very
quickly with a quantum version of the well-known Fast Fourier
Transform algorithm (FFT). Actually applying the QFFT to a $n$ qubit
register takes only $O(n^2)$ quantum gates, or even less. Of course,
these have to be ``non-classical'' gates (transforming ``computational
basis states'' into superpositions). Without going into details, just
note that two types of gates are used: the 1-qubit ``Hadamard
transform''

$$H: \quad \ket{0} \to \frac{1}{\sqrt{2}} (\ket{0}+\ket{1}) \qquad
\ket{1} \to \frac{1}{\sqrt{2}} (\ket{0}-\ket{1})$$ 
and 2-qubit ``controlled phase shift" gates which do $\ket{11} \to
e^{i \phi} \ket{11}$, and leave the other 3 basis states unchanged.

Fourier transforming a periodic function will result in equally spaced
peaks with no offset (thus, the first peak is at 0). In particular
when transforming our equally spaced peaks we again get peaks and the
offset will only show up in the phases of the resulting amplitudes. (A
minor complication is that the peaks are now a bit ``smeared
out". This is because in general the size $2^n$ of the Fourier
transformation is not a multiple of the period $r$.)

Below are pictures of the probability (amplitude absolute value
squared) for the 64 basis states of a 6 qubit register before and
after the QFFT. (For the first picture, we could also say that it
shows the amplitudes themselves, as they are anyway real and
positive.)

\pagebreak

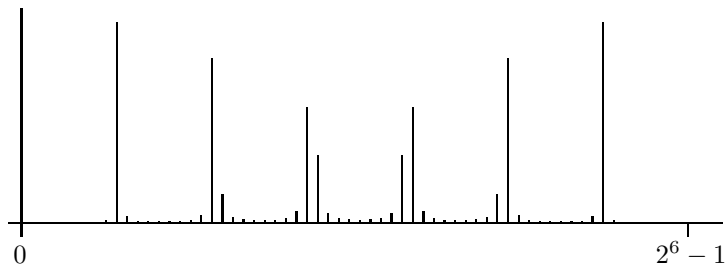
\begin{figure}
\begin{center}
\begin{picture}(300,150)

\thinlines \put(-5,0){\line(1,0){270}}

\put(0,0){\line(0,-1){5}} \put(-3,-15){0}
\put(252,0){\line(0,-1){5}} \put(240,-15){$2^6-1$}

\put(0,0){\line(0,1){81.0}}
\put(4,0){\line(0,1){0.0}}
\put(8,0){\line(0,1){0.0}}
\put(12,0){\line(0,1){0.0}}
\put(16,0){\line(0,1){0.0}}
\put(20,0){\line(0,1){0.1}}
\put(24,0){\line(0,1){0.1}}
\put(28,0){\line(0,1){0.3}}
\put(32,0){\line(0,1){1.0}}
\put(36,0){\line(0,1){75.9}}
\put(40,0){\line(0,1){2.6}}
\put(44,0){\line(0,1){0.7}}
\put(48,0){\line(0,1){0.4}}
\put(52,0){\line(0,1){0.4}}
\put(56,0){\line(0,1){0.4}}
\put(60,0){\line(0,1){0.6}}
\put(64,0){\line(0,1){1.0}}
\put(68,0){\line(0,1){3.0}}
\put(72,0){\line(0,1){62.2}}
\put(76,0){\line(0,1){10.9}}
\put(80,0){\line(0,1){2.2}}
\put(84,0){\line(0,1){1.1}}
\put(88,0){\line(0,1){0.8}}
\put(92,0){\line(0,1){0.8}}
\put(96,0){\line(0,1){1.0}}
\put(100,0){\line(0,1){1.6}}
\put(104,0){\line(0,1){4.1}}
\put(108,0){\line(0,1){43.7}}
\put(112,0){\line(0,1){25.3}}
\put(116,0){\line(0,1){3.7}}
\put(120,0){\line(0,1){1.7}}
\put(124,0){\line(0,1){1.1}}
\put(128,0){\line(0,1){1.0}}
\put(132,0){\line(0,1){1.1}}
\put(136,0){\line(0,1){1.7}}
\put(140,0){\line(0,1){3.7}}
\put(144,0){\line(0,1){25.3}}
\put(148,0){\line(0,1){43.7}}
\put(152,0){\line(0,1){4.1}}
\put(156,0){\line(0,1){1.6}}
\put(160,0){\line(0,1){1.0}}
\put(164,0){\line(0,1){0.8}}
\put(168,0){\line(0,1){0.8}}
\put(172,0){\line(0,1){1.1}}
\put(176,0){\line(0,1){2.2}}
\put(180,0){\line(0,1){10.9}}
\put(184,0){\line(0,1){62.2}}
\put(188,0){\line(0,1){3.0}}
\put(192,0){\line(0,1){1.0}}
\put(196,0){\line(0,1){0.6}}
\put(200,0){\line(0,1){0.4}}
\put(204,0){\line(0,1){0.4}}
\put(208,0){\line(0,1){0.4}}
\put(212,0){\line(0,1){0.7}}
\put(216,0){\line(0,1){2.6}}
\put(220,0){\line(0,1){75.9}}
\put(224,0){\line(0,1){1.0}}
\put(228,0){\line(0,1){0.3}}
\put(232,0){\line(0,1){0.1}}
\put(236,0){\line(0,1){0.1}}
\put(240,0){\line(0,1){0.0}}
\put(244,0){\line(0,1){0.0}}
\put(248,0){\line(0,1){0.0}}
\put(252,0){\line(0,1){0.0}}

\end{picture}
\end{center}
\vspace*{1cm}
\caption{Probabilities (amplitudes absolute value squared) of
observing each basis state after the QFFT. Now the peaks are not
``sharp" any more. Approximately they are spaced by $2^n/r$, thus here
$64/7 \approx 9.14$.}
\label{QFFT}
\end{figure}

\vspace{1.5cm}

{\bf exercise:}(maybe a bit cumbersome) Check that the discrete
Fourier transform qualitatively really does what's indicated in the
picture above: The resulting peaks start at 0 and are spaced by
$2^n/r$. Check that the width of the resulting peaks is of order
$O(1)$ for $2^n \approx r^2$ or larger. (In Shor's algorithm $n$ is
chosen such that $2^n \approx (pq)^2$ or larger.)

(Note: the QFFT Fourier transforms $2^n$ amplitudes in only some
$O(n^2)$ steps, while a conventional FFT would take $O(2^n \log 2^n)$
steps to transform $2^n$ numbers. Still, we cannot say that a quantum
computer can Fourier transform much faster than a classical computer,
as the operations are not comparable. Fourier transforming the
amplitudes of a register has no classical analogue and cannot be used
to do the job of a classical FFT.)

After the QFFT, we measure the register. Thus with high probability we
will measure an integer close to $k \cdot 2^n/r$ for some random
integer $k$. By dividing the observed integer by $2^n$ we get a
fraction close to $k/r$. Then we can use the continued fraction
expansion of this fraction, which gives a sequence of best rational
approximations. With luck a single run will be enough to find $k$
(which doesn't interest us) and $r$. Otherwise combining the output of
several runs of the algorithm (different $k$'s) will allow to find $r$
with high probability.

As mentioned above, measuring the output register (which holds $f(x)
$) was not necessary. After all we didn't use the measured value
$f(x_0)$.

What would have happened if we had not uncomputed the ``garbage''?
Then the computed function would have been the pair $f(x),g(x)$ and it
would not have been periodical, thus the algorithm wouldn't have
worked.

\section{Computer science aspects and further remarks}

\subsection{On the ``power'' of quantum computation}
In what respect is a quantum computer thought to be more powerful than
a conventional computer? In computer science, people typically ask
about the \underline{scaling} of the computation time (number of
elementary operations) with the size of a problem (number of bits of
the input). So multiplying two $n$-digit numbers normally takes
$O(n^2)$ gates (it can also be done a bit faster), while in classical
factoring algorithms the computation time grows faster than any power
of the length (number of digits) of the number to be factored.

In ``computational complexity theory'' people are interested in how
the best possible algorithm for a problem scales. The main distinction
is between problems that can be solved in ``polynomial time'' (time
grows at most like some power of the input size) and those which take
longer, in particular exponential time. It is believed that many
interesting problems like factoring cannot be solved in polynomial
time, although so far it hasn't been proven for any natural
problem. Shor's algorithm factors in polynomial time (actually
$O(n^3)$ because of modular exponentiation), so if classically
factoring really is hard (can't be done in polynomial time), it
follows that quantum computers are more ``powerful'' than conventional
ones.

\subsection{Trying to solve NP-complete problems}
Problems of the type NP (NP stands for ``nondeterministic
polynomial'') are roughly those where, once one has an answer, one can
quickly check whether it really is correct. More formally, in NP
problems, the answer is just ``yes/no'', and if it is ``yes'', there
exists a short (polynomial time) proof that it really is ``yes''. Most
natural such problems are related, in that any one can be reduced to
any other one. These problems are called ``NP-complete''. Thus if one
finds an algorithm that can solve such a problem (in polynomial time),
one can solve all of them (actually all NP problems). The question
whether there is a conventional, efficient (polynomial time) such
algorithm is the famous ``NP=P(?)'' question. (It is widely believed
that NP$\not=$P, but it's still unproven...)

Factoring is not NP-complete, but it is NP (checking whether a given
number really is a factor is easy). It would be a big breakthrough if
somebody would manage to find an efficient (polynomial time) quantum
algorithm for an NP-complete problem. But it seems not to be easy and
may well not be possible at all.

\subsection{Subtle questions about scaling of precision; fault tolerance}
When talking about what kind of computers are possible, questions
about the precision of their parts and whether this precision has to
increase for longer computations should not be forgotten. For
classical computers, digital (electronic) hardware seems not to have
much of a scaling problem.

However, quantum computers seem to have a problem similar to that of
classical analogue computers: There is a continuum of states and
gates. If we shine a laser a bit too long at an ion, there is no
automatic resetting like in digital electronics. Rather, the errors in
the amplitudes will grow larger and larger as we do more gates. Also
the very non-classical state of the quantum computer is sensitive to
decoherence. Thus, it seems that the longer a quantum computation is,
the more precise the gates would have to be.

Fortunately, it has been shown that clever error-correcting techniques
are possible which make quantum computations ``fault tolerant''. In
these techniques, only a subspace of the whole QC is used. Qubits are
encoded in complicated entangled subspaces of several physical qubits,
and gates are done by really doing a whole sequence of gates on the
physical qubits.

Fault tolerant quantum computing is maybe the second most important
development after Shor's algorithm in this field. Without it, running
interesting quantum algorithms would be only a very distant
possibility.

\subsubsection{Is entanglement necessary for quantum computation?}
One could argue that a quantum computation could be carried out with
any high dimensional quantum system. E.g., instead of using many
qubits one could use the energy eigenstates of a hydrogen
atom. Because this would be a single system, there would be no
entanglement. The problem is that in such systems the precision
problem scales much worse and effectively destroys the (possible)
advantage of quantum computing.

Therefore, the reason why entanglement is necessary for ``real''
quantum computing is because when we talk about the power of a
computing model we talk about scaling issues, and precision is an
important aspect of this.

So what went wrong when computer scientists thought that their formal
classical computing models (in particular the Turing machine) captured
the full computational power of nature? One can argue that the reason
is, that they didn't think of the possibility of entangled states,
thus that the state of a system cannot be described by describing the
state of each part separately.

Maybe this is not the full truth: A different aspect of the question
whether entanglement is necessary has arisen from thinking about
quantum computation with highly mixed states, as in NMR (nuclear
magnetic resonance) quantum computing experiments. If the state is
close enough to the maximally mixed state, it necessarily is
separable, thus it can be viewed as a probabilistic mixture of product
states. Still it is not clear whether such an ``unentangled quantum
computation" can efficiently be simulated with probabilistic classical
computation. The problem is how to efficiently arrive from the
statistical ensemble before a quantum gate to the one after the
quantum gate. Although none of the major quantum algorithms (Shor's,
Grover's...) seem to work on such a quantum computer, it still may
outperform conventional computers on some other problems.

\subsection{Other quantum algorithms}
There are a number of other quantum algorithms besides Shor's, but
many of them solve rather ``academic'' and somewhat unnatural problems
which are primarily of interest to computer scientists. Often
something has to be found out about ``black box'' functions, thus we
can only evaluate the function, but we do not see how it is
calculated. Grover's, Simon's and the Deutsch-Jozsa algorithm are all
of this kind.

\subsubsection{Grover's ``quantum searching" algorithm}
The problem that Grover's algorithm solves can be described like this:
Say we have a classical algorithm which searches through a big number
$N$ of cases (e.g. numbers) to find the single one which fulfills some
criteria. Then Grover's algorithm can find such an object in only
about $\sqrt{N}$ instead of $N$ search steps. One could e.g. search
through all possible next 10 moves of a chess player. However, usually
there are better ways to find solutions than to simply search through
all possibilities, so Grover's algorithm may not be so useful.

Also one can solve any NP problem by searching through exponentially
many cases, but then Grover's algorithm would still take exponentially
many steps, so it doesn't provide a fast solution. Also it has been
shown that such ``simple'' unstructured searches can't be done faster
than by Grover's algorithm. Thus this ``black box" approach to
efficiently solve NP problems does not work.

\subsubsection{communication complexity}
Also many algorithms solve ``communication complexity'' problems,
where e.g. the input is split between two distant places and the
problem should be solved with as little communication as
possible. There are such algorithms where exponentially fewer qubits
have to be sent that the number of bits that would have to be sent
classically.

\subsection{Are even more powerful computers possible?}
First: can quantum computers even work? In a way, a working quantum
computer would test quantum theory in a way that has not been done so
far. In particular the ``existence'' of the large Hilbert spaces of
systems with many parts would be tested. A quantum computer could more
or less at random produce states in this space and check whether they
behave as expected. Thus, the question is whether quantum theory
really is correct or whether it is only an ``effective'' theory that
works for not too high dimensional Hilbert spaces.

Probably quantum theory is correct. Very few physicists would doubt
that at least the basic framework (superposition principle, etc.) is
correct. Still, a working QC would provide a very strong test. Also
theoretical quantum computing could be of interest to quantum
physics. In a way, the complicated quantum formalism with the huge
information content of states is only a tool to arrive at relatively
simple results. If it could be shown that quantum computers are more
powerful than classical ones, it would be clear that this formalism
could not be simplified a lot while still producing the same results.

Could there be computers that are more powerful? So far, there is no
indication of this in physics, but it is at least imaginable. E.g., it
is known that more powerful quantum algorithms would exist if quantum
theoretic time evolution were not exactly linear. This would allow to
efficiently solve NP-complete problems and even more, but it seems
improbable that such nonlinearities exist. \\

\paragraph{recommended literature:\\}
Shor's original 26 page article \cite{Shor} is a good entry
point. \\


\begin{thebibliography}{10}

\bibitem{Shor}
P. Shor, {\it Polynominal-Time Algorithms for Prime Factorization and Discrete
Logarithms on a Quantum Computer.}
In {\it Proc. 35th Annual Symposium on Foundations of Computer
Science.} IEEE Press, pp 124-134, Nov. 1994, quant-ph/9508027

\bibitem{grover}
L. Grover, {\it A fast quantum mechanical algorithm for database search},
Proceedings, 28th Annual ACM
Symposium on the Theory of Computing (STOC), May 1996, pages 212-219,
also quant-ph/9605043

\bibitem{shor3}
P. Shor,
{\it Scheme for reducing decoherence in quantum computer memory}
Phys. Rev. A {\bf 52}, No. 4, pp. R2493-R2496 (1995)

\bibitem{css}
A.R. Calderbank and P. Shor,
{\it Good Quantum Error-Correcting Codes Exist},
Phys. Rev. A {\bf 54}, No. 2, 1098-1106 (1996),
also quant-ph/9512032

\bibitem{shor2}
P. Shor,
{\it Fault-tolerant quantum computation},
37th Symposium on Foundations of Computing, IEEE Computer Society
Press, 1996, pp. 56-65,
also quant-ph/9605011 

\bibitem{cirac}
J.I. Cirac and P. Zoller,
{\it Quantum Computations with Cold Trapped Ions},
Phys. Rev. Lett. {\bf 74}, No. 200, 4091 (May 1995)

\end{thebibliography}
\end{document}